\input harvmac.tex
\input epsf
\noblackbox

\newcount\figno

\figno=0
\def\fig#1#2#3{
\par\begingroup\parindent=0pt\leftskip=1cm\rightskip=1cm\parindent=0pt
\baselineskip=11pt \global\advance\figno by 1 \midinsert
\epsfxsize=#3 \centerline{\epsfbox{#2}} \vskip 12pt
\centerline{{\bf Figure \the\figno :}{\it ~~ #1}}\par
\endinsert\endgroup\par}
\def\figlabel#1{\xdef#1{\the\figno}}


\def\inbar{\vrule height1.5ex width.4pt depth0pt}
\def\IC{\relax\,\hbox{$\inbar\kern-.3em{\rm C}$}}
\def\IR{\relax{\rm I\kern-.18em R}}
\def\IN{\relax{\rm I\kern-.18em N}}
\def\IP{\relax{\rm I\kern-.18em P}}
\def\ZZ{\bb{Z}}
\def\narrowplus{\kern -.04truein + \kern -.03truein}
\def\narrowminus{- \kern -.04truein}
\def\narrowminussub{\kern -.02truein - \kern -.01truein}

\font\mbm = msbm10
\def\bb#1{\hbox{\mbm #1}}

\lref\antmass{I. Antoniadis, K.S. Narain, T.R. Taylor, hep-th/0507244.}
 \lref\pran{A.H.
Chamseddine, R. Arnowitt and P. Nath, Phys. Rev. Lett. 49 (1982) 970; R. Barbieri, S.
Ferrara and C.A. Savoy, Phys. Lett. B119 (1982) 343.}
 \lref\atick{J.J. Atick,
L.J. Dixon and P.A. Griffin, Nucl.\ Phys.\ B 298 (1988) 1.} \lref\fay{John D. Fay, {\it
Theta Functions on Riemann Surfaces}, in {\it Lecture Notes in Mathematics}
(Springer-Verlag, 1973).}
\lref\mumford{David Mumford, {\it Tata Lectures on Theta II} (Birkh\"auser,
1993).}
\lref\ver{E. Verlinde and H. Verlinde, Nucl.\ Phys.\ B 288 (1987) 357.}
\lref\dhok{E. D'Hoker and D.H. Phong,
Nucl.\ Phys.\ B 639 (2002) 129, hep-th/0111040.}
\lref\narain{K.~S.~Narain, M.~H.~Sarmadi and C.~Vafa,
Nucl.\ Phys.\ B 356 (1991) 163.}
\lref\morozov{A.~A.~Belavin, V.~Knizhnik, A.~Morozov and A.~Perelomov,
JETP Lett.\  43 (1986) 411
[Phys.\ Lett.\ B 177 (1986) 324].}
\lref\moore{G. Moore, Phys.\ Lett.\ B 176 (1986) 369.}
\lref\SS{J.~Scherk and J.~H.~Schwarz,
Phys.\ Lett.\ B 82 (1979) 60.}
\lref\rt{R.~Rohm,
Nucl.\ Phys.\ B 237 (1984) 553; H. Itoyama and T.R. Taylor, Phys.\
Lett.\ B 186 (1987) 129.} \lref\kpor{C.~Kounnas and M.~Porrati,
Nucl.\ Phys.\ B 310 (1988) 355.}
 \lref\SSII{
S.~Ferrara, C.~Kounnas, M.~Porrati and F.~Zwirner,
Nucl.\ Phys.\ B 318 (1989) 75;
C.~Kounnas and B.~Rostand,
Nucl.\ Phys.\ B 341 (1990) 641;
I.~Antoniadis,
Phys.\ Lett.\ B 246 (1990)~377.}
\lref\SSI{I.~Antoniadis, E.~Dudas and A.~Sagnotti,
Nucl.\ Phys.\ B 544 (1999) 469 [arXiv:hep-th/9807011].}
\lref\ABLM{
I.~Antoniadis, K.~Benakli, A.~Laugier and T.~Maillard,
Nucl.\ Phys.\ B 662 (2003) 40, hep-ph/0211409.} 
\lref\invo{ S.~K.~Blau, S.~Carlip,
M.~Clements, S.~Della Pietra and V.~Della Pietra, Nucl.\ Phys.\ B 301 (1988) 285; 
M.~Bianchi and A.~Sagnotti, Phys.\ Lett.\ B 211 (1988) 407.} 
\lref\modular{ M.~Bianchi and A.~Sagnotti, Phys.\ Lett.\ B 231 (1989) 389.} 
\lref\at{I.~Antoniadis and T.~Taylor, Nucl.\ Phys.\ B 695
(2004) 103 [arXiv:hep-th/0403293].} 
\lref\ant{ I.~Antoniadis, K.~S.~Narain and T.~Taylor, in preparation.} 
\lref\Oog{ H.~Ooguri and C.~Vafa, hep-th/0302109.} 
\lref\bcov{M. Bershadsky,
S. Cecotti, H. Ooguri and C. Vafa, Commun. Math Phys. 165 (1994) 311, hep-th/9309140.}
\lref\agnt{ I.~Antoniadis, E.~Gava, K.~S.~Narain and T.~Taylor, Nucl.\ Phys.\ B 413 (1994)
162, hep-th/9307158.} 
\lref\aq{ I.~Antoniadis and M.~Quiros,
Nucl.\ Phys.\ B {505} (1997) 109, hep-th/9705037;
R.~Rattazzi, C.~A.~Scrucca and A.~Strumia,
Nucl.\ Phys.\ B {674} (2003) 171, hep-th/0305184.}
\lref\rs{
L.~Randall and R.~Sundrum,
Nucl.\ Phys.\ B {557} (1999) 79 [arXiv:hep-th/9810155]; G. Giudice, M.A. Luty,
H. Murayama and R. Rattazzi, JHEP 9812 (1998) 027 [arXiv:hep-ph/9810442].}
\lref\dfms{
L.~Dixon, D.~Friedan, M.~Martinec and S.~Shenker,
Nucl.\ Phys.\ B {282} (1987) 13.}
\lref\bsb{
S.~Sugimoto,
Prog.\ Theor.\ Phys.\ 102 (1999) 685, hep-th/9905159;
I.~Antoniadis, E.~Dudas and A.~Sagnotti,
Phys.\ Lett.\ B {\bf 464} (1999) 38, hep-th/9908023;
G.~Aldazabal and A.~M.~Uranga,
JHEP {\bf 9910} (1999) 024, hep-th/9908072.}
\lref\nl{
E.~Dudas and J.~Mourad,
Phys.\ Lett.\ B {\bf 514} (2001) 173, hep-th/0012071;
G.~Pradisi and F.~Riccioni,
Nucl.\ Phys.\ B {\bf 615} (2001) 33, hep-th/0107090.}
\lref\cutoff{
I.~Antoniadis,
Phys.\ Lett.\ B  246 (1990) 377;
I.~Antoniadis, S.~Dimopoulos and G.~R.~Dvali,
Nucl.\ Phys.\ B  516 (1998) 70
[arXiv:hep-ph/9710204];
I.~Antoniadis, S.~Dimopoulos, A.~Pomarol and M.~Quiros,
Nucl.\ Phys.\ B 544 (1999) 503
[arXiv:hep-ph/9810410].}
\lref\mgaugino{
I.~Antoniadis and M.~Quiros,
Nucl.\ Phys.\ B 505 (1997) 109
[arXiv:hep-th/9705037];
M.~A.~Luty and N.~Okada,
JHEP 0304 (2003) 050
[arXiv:hep-th/0209178];
R.~Rattazzi, C.~A.~Scrucca and A.~Strumia,
Nucl.\ Phys.\ B 674 (2003) 171
[arXiv:hep-th/0305184].}

\Title{\vbox{
     \hbox{CERN--PH--TH/2005-155}
    \hbox{hep-th/0509048}}}
{\vbox{\centerline{Note on Mediation of Supersymmetry Breaking}\vskip 3mm\centerline{ from
Closed to Open Strings}}}
\centerline{Ignatios Antoniadis{$^{1,\dagger}$} and Tomasz
R.\ Taylor{$^{2}$ }}
\bigskip\medskip
\centerline{$^1${\it CERN Theory Division, CH-1211
Geneva 23, Switzerland}}
\centerline{$^2${\it Department of Physics, Northeastern University, Boston, MA
02115, U.S.A.}}
\bigskip
\bigskip

\centerline{\bf Abstract} \noindent 
We discuss the mediation of supersymmetry breaking from closed to open strings,
extending and improving previous analysis of the authors in Nucl.\ Phys.\ B 695 (2004) 103 [hep-th/0403293].
In the general case, we find the absence of anomaly mediation around any 
perturbative string vacuum.
When supersymmetry is broken by Scherk-Schwarz boundary conditions along
a compactification interval
perpendicular to a stack of D-branes, the gaugino acquires a mass 
at two loops that behaves as $m_{1/2}\sim g^4 m_{3/2}^3$ in string units, where
$m_{3/2}$ is the gravitino mass and $g$ is the gauge coupling. 

\vfill\hrule\vskip 1mm\noindent {$^{\dagger}$ 
On leave from
CPHT Ecole Polytechnique (UMR du CNRS 7644) F-91128, Palaiseau.} 
\Date{}
The purpose of this note is to extend the previous discussion {\at} of mediation of
supersymmetry breaking between closed and open string sectors. At the same time, we will
also improve and clarify some computations presented in \at.

The mediation of supersymmetry breaking is (by definition) the mechanism responsible for
communicating supersymmetry breaking from a hidden sector to the spectrum of observable
particles. Although there are several possible ways how such a mechanism may be realized in
low-energy effective field theory, most of its concrete implementations involve
non-renormalizable interactions, therefore the outcome can be sensitive to the ultraviolet
completion of the theory. Hence the finite superstring theory offers a valuable framework
for studying the effects of supersymmetry breaking while keeping the ultra-violet physics
under control. 

One type of mediation possible in this context, is the so-called anomaly
mediation \rs, providing a contribution to the gaugino mass that scales linearly with
the gravitino mass: $m_{1/2}\sim b_0 g^2 m_{3/2}$, where $g$ is  the gauge
coupling and  $b_0$ is the coefficient of
the corresponding one-loop beta-function.
However, as explained in \at, this contribution is absent in any perturbative 
string vacuum. The reason is that such a result should arise at one-loop
level, as dictated by the power of the gauge coupling, {\it e.g.} on a world-sheet
with two boundaries (annulus) or one boundary and a crosscap (M\"obius strip).
The corresponding string diagram involving two left-handed gauginos at zero
momentum vanishes though, due to the $U(1)$ charge conservation of the
two-dimensional (2d) $N=2$ superconformal symmetry. Indeed, the massless 
gaugino vertex operator of definite chirality $\alpha$, at zero momentum, 
in the canonical $-1/2$-ghost picture, reads:
\eqn\vertex{
V^{(-1/2)}_\alpha(x)=:e^{-\varphi/2}S_\alpha e^{i{\sqrt 3\over 2}H}:
\, ,}
where $x$ is a position on the boundary of the world-sheet, $\varphi$ is
the scalar bosonizing the superghost system, $H$ is the free 2d boson
associated to the $N=2$ $U(1)$ current $J=i{\sqrt 3}\partial H$, and we
neglected the Chan-Paton gauge indices for simplicity. The two-point function
involves, besides the two gauginos of the same chirality at the boundary 
of the world-sheet (annulus or M\"obius strip), one picture changing 
operator (PCO). The latter can provide at most $-1$ charge which is not
sufficient for cancelling the $+3$ $U(1)$ charge of the gauginos and
thus, the amplitude vanishes. Charge cancellation can be achieved at
higher order, requiring a Riemann surface of Euler characteristic at least 
equal to $-1$. An example of such a surface contains one handle and
one boundary and will be studied in the example described below.

The particular setup considered in \at\ is the gravitational mediation from
the closed string sector with supersymmetry broken by Scherk-Schwarz \SS\ boundary
conditions in one of the compact directions, which plays the role of the hidden sector, to
the ``observable'' sector of open strings ending on D-branes perpendicular to the
Scherk-Schwarz direction \SSI. The corresponding gauginos that remain massless at the tree level,
acquire masses due to the world-sheet diagram with one handle and one boundary -- genus
$g=1$ Riemann surface $\Sigma$ with $h=1$ boundary, {\it i.e.} with Euler characteristic
${-}1$, see Fig.1. \fig{Bordered $g=1$ surface $\Sigma$ with the two gaugino vertices
inserted at the boundary}{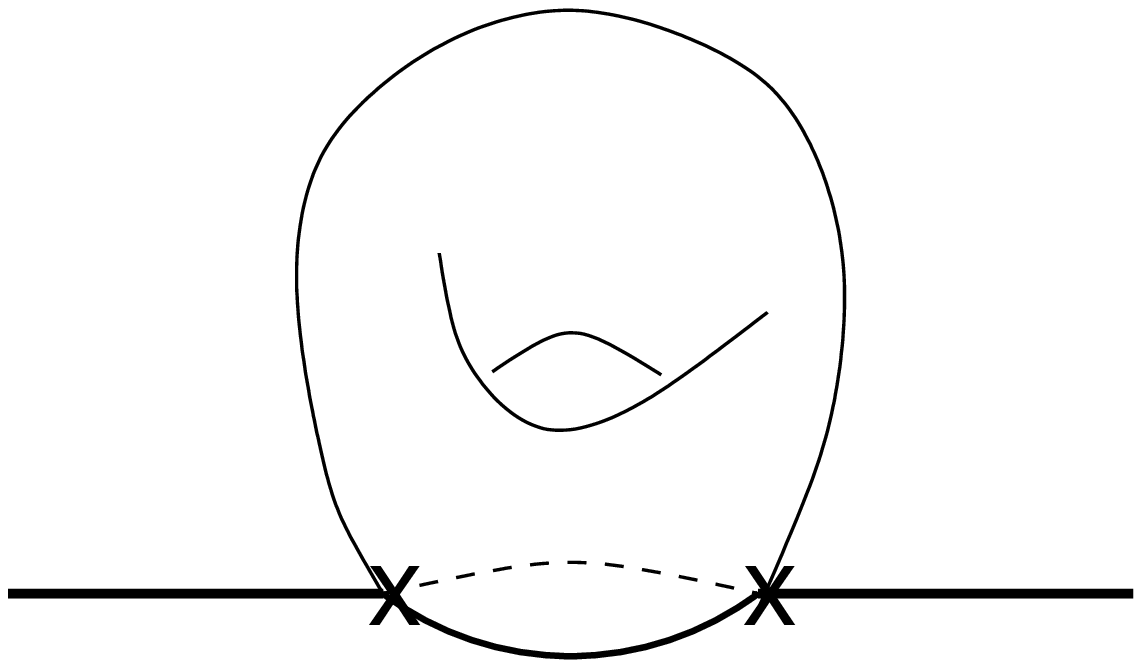}{4cm}

In Ref.\at, we discussed type II compactifications on $T^2\times K3$, with the
Scherk-Schwarz circle of radius $R$ embedded in $T^2$. Although the generic case can be
studied to some extent, the mass computations simplify enormously in the orbifold limit,
with K3 represented as the quotients $T^4/\ZZ_2$ or $T^4/\ZZ_N$. Unfortunately, in this case
the mass is protected by the orbifold symmetries -- the remnants of the continuous
internal rotations, or equivalently
R-symmetries of the low-energy effective field theory -- hence the result is zero. We will
first review the origin of this result and then we will try to circumvent it by  blowing up
the orbifold singularity of $K3$.

The gaugino mass is given by the following integral \at: 
\eqn\mass{
m_{1/2}=g_s^2\int_{F(\Sigma)} d\mu\int_{\partial\Sigma}
      dxdy\ {\cal A}(x,y),}
where $d\mu$ is the measure of the moduli space of $\Sigma$, with the integration extending
over the fundamental domain ${F(\Sigma)}$, while the two-point function 
\eqn\amp{\eqalign{
{\cal A}(x,y)= ~&\theta\left[{0\atop {1\over 2}\vec 1}\right](x-\Delta)
{\sigma(x)\sigma(y)\over\prod_{I<J}^{3,4,5}E(z_I,z_J) \prod_{I=3}^5\sigma^2(z_I)}\cr ~&
\times\prod_{I=3}^5 \theta_{h_I^{-1}}\left[{0\atop {1\over 2}\vec 1}
\right](z_I-\Delta)\,\partial X_{h_I}(z_I)\,{\cal Z} .}} 
is integrated over the boundary. In the mass formula \mass, $g_s=g^2$ is the string coupling.
The additional points, $z_I$, labeled by the internal planes $I=3,4,5$, take values in the
set of the insertion points, $z_a$, $a=1,2,3$, of PCOs, and
the above expression should be summed over all permutations $\{z_{I(a)}\}$. Although {\it a
priori} arbitrary, the PCO insertion points are subject to the constraint
\eqn\gauged{\sum_{I=3}^{I=5}z_I=y+2\Delta\, .} 
as a result of a gauge choice made in arriving to \amp. 
Here, $\Delta$ is the Riemann $\theta$-constant. 
Although the above gauge choice is  formally not allowed \antmass, it can be 
used throughout the computations.
Indeed, by inserting another vertex of an open string Wilson line, one can use
an appropriate gauge condition and show that the amplitude can be written 
as the variation with respect to the Wilson line of the original one,
evaluated by formally using the condition \gauged.

After summing over all 6
permutations, the two-point function \amp\ should become manifestly independent of the PCO
insertion points. In Eq.\amp, $\partial X_{h_I}$ are the zero modes (instanton
contributions) twisted by the orbifold group elements $h_I$ while the position-independent
factor $\cal Z$ includes the lattice partition function as well as all non-zero mode
determinants. Finally, $\sigma$ is the one-differential with no zeroes or poles and $E$ is
the prime form. The crucial property of the prime form is the antisymmetry
$E(z_I,z_J)=-E(z_J,z_I)$. As a result, the permutation sum amounts to antisymmetrizing the
factor
\eqn\fact{\theta_{h_3^{-1}}\left[{0\atop {1\over 2}\vec 1} \right](z_3-\Delta)\,\partial
X_{h_3}(z_3)\;K(z_4,z_5),\qquad K(z_4,z_5)=\prod_{I=4}^5 \theta_{h_I^{-1}}\left[{0\atop
{1\over 2}\vec 1} \right](z_I-\Delta)\,\partial X_{h_I}(z_I)} 
in the positions $z_a$. Note
that, for the $T^2\times K_3$ compactification under consideration, $h_3=1$ while $h_4=h$,
$h_5=h^{-1}$, where $h$ is the element of the  $K3$ orbifold group $\ZZ_N$. Clearly, in the
case of $\ZZ_2$, $h=h^{-1}$, hence the result vanishes upon antisymmetrization. As shown in
\at, similar conclusion holds for arbitrary $\ZZ_N$, at least up to terms that are
exponentially suppressed in the large Scherk-Schwarz radius limit. As announced before, we
will try to avoid this conclusion by blowing up the orbifold singularity. This can be
achieved by switching on the vacuum expectation value of one of the blowing-up modes. Thus
the amplitude will now include an additional insertion of the vertex operator creating the
twisted blowing-up mode $\cal B$ at zero momentum. Note that this additional vertex must be
inserted in the $0$-ghost picture in order to preserve the balance of the background ghost
charge.

In the $-1$-ghost picture, the zero momentum vertex of a blowing-up mode $\cal B$ associated
to the twisted sector $(h,h^{-1})$, with $h=e^{2i\pi\epsilon}$ and $\epsilon=k/N$, reads:
\eqn\twistvertex{ V^{(-1,-1)}_{\cal B}(\zeta,\bar\zeta)=:e^{-i\epsilon H_4}e^{-i(1-\epsilon)
{\tilde H}_4} \sigma_{--}^{4,1-\epsilon}e^{-i(1-\epsilon) H_5}e^{-i\epsilon {\tilde
H}_5}\sigma_{--}^{5,\epsilon}:\, ,} 
where $\sigma_{--}$ is the corresponding twist field of
conformal dimension $\epsilon(1-\epsilon)/2$ in both left and right movers. Here, $(H_4,
H_5)$ and $({\tilde H}_4, {\tilde H}_5)$ are the scalars bosonizing the left- and
right-moving fermionic coordinates of $K3$, respectively.\foot{Since the following paragraph
corrects several (accumulative) misprints contained in \at, we deliberately change the
notation and repeat the computation of the vertex operator from the scratch.} In order to
change the picture, we use the supercurrents \eqn\superc{T_L=\sum_{I=4,5}(\partial
X^Ie^{-iH_I}+\partial {\bar X}^Ie^{iH_I}) \qquad T_R=\sum_{I=4,5}(\bar\partial
X^Ie^{-i{\tilde H}_I} +\bar\partial {\bar X}^Ie^{i{\tilde H}_I}).} We use the OPE rules
\dfms \eqn\oper{\eqalign{&\sigma_{--}^{I,\epsilon}(z,\bar z)
\partial {\bar X}_I(\bar w)\sim
( z-w)^{-\epsilon}\sigma_{+-}^{I,\epsilon}(z,\bar z)\qquad
\sigma_{--}^{I,\epsilon}(z,\bar z)    \bar\partial {\bar X}_I(\bar w)\sim
(\bar z-\bar w)^{-1+\epsilon}\sigma_{-+}^{I,\epsilon}(z,\bar z)\cr &
\sigma_{--}^{I,\epsilon}(z,\bar z)\partial
X_I(w)\sim (z-w)^{-1+\epsilon}\sigma_{+-}^{I,\epsilon}(z,\bar z)\qquad
\sigma_{--}^{I,\epsilon}(z,\bar z)\bar\partial
X_I(w)\sim (\bar z-\bar w)^{-\epsilon}\sigma_{-+}^{I,\epsilon}(z,\bar z)
}}
with the further short-distance expansion
\eqn\opera{\eqalign{&\sigma_{-+}^{I,\epsilon}(z,\bar z)
\partial {\bar X}_I(\bar w)\sim
( z-w)^{-\epsilon}\sigma_{++}^{I,\epsilon}(z,\bar z)\qquad \sigma_{+-}^{I,\epsilon}(z,\bar
z)    \bar\partial {\bar X}_I(\bar w)\sim (\bar z-\bar
w)^{-1+\epsilon}\sigma_{++}^{I,\epsilon}(z,\bar z)\cr & \sigma_{-+}^{I,\epsilon}(z,\bar
z)\partial X_I(w)\sim (z-w)^{-1+\epsilon}\sigma_{++}^{I,\epsilon}(z,\bar z)\qquad
\sigma_{+-}^{I,\epsilon}(z,\bar z)\bar\partial X_I(w)\sim (\bar z-\bar
w)^{-\epsilon}\sigma_{++}^{I,\epsilon}(z,\bar z) }} Using the $N=2$ world-sheet
supercurrent, one finds the blowing-up vertex operator in the 0-ghost picture:
\eqn\twistvertexone{\eqalign{ V^{(0,0)}_{\cal B}(\zeta,\bar\zeta)=&
:\sigma_{+-}^{4,1-\epsilon}e^{i(1-\epsilon) H_4}e^{-i(1-\epsilon) {\tilde
H}_4}\;\sigma_{-+}^{5,\epsilon} e^{-i(1-\epsilon) H_5}e^{i(1-\epsilon) {\tilde H}_5} \cr
&+\sigma_{-+}^{4,1-\epsilon}e^{-i\epsilon H_4}e^{i\epsilon {\tilde
H}_4}\;\sigma_{+-}^{5,\epsilon} e^{i\epsilon H_5}e^{-i\epsilon {\tilde H}_5}\cr &
+\sigma_{++}^{4,1-\epsilon}e^{i(1-\epsilon) H_4}e^{i\epsilon {\tilde
H}_4}\;\sigma_{--}^{5,\epsilon} e^{-i(1-\epsilon) H_5}e^{-i\epsilon {\tilde H}_5}\cr &+
\sigma_{--}^{4,1-\epsilon}e^{-i\epsilon H_4}e^{-i(1-\epsilon) {\tilde
H}_4}\;\sigma_{++}^{5,\epsilon} e^{i\epsilon H_5}e^{i(1-\epsilon) {\tilde H}_5}: }}

The modification of the amplitude due to the the insertion of the blowing-up vertex operator
$\int d^2\zeta V^{(0,0)}_{\cal B}(\zeta,\bar\zeta)$ can be obtained by repeating step by
step the derivation of \amp\ presented in \at. The only change is in the $K3$ part of the
amplitude where, for a given spin structure $s$, the following correlators appear:
\eqn\Kthree{\eqalign{&\left\langle \sigma_{+-}^{4,1-\epsilon}(\zeta,\bar\zeta)\partial
X^4(z_4) \right\rangle\left\langle e^{iH_4/2}(x)e^{iH_4/2}(y)e^{-iH_4}(z_4) e^{i(1-\epsilon)
H_4}(\zeta) e^{-i(1-\epsilon) {\tilde H}_4}(\bar\zeta)\right\rangle_s\times \cr &
\left\langle \sigma_{-+}^{5,\epsilon}(\zeta,\bar\zeta)\partial X^5(z_5)\right\rangle
\left\langle e^{iH_5/2}(x)e^{iH_5/2}(y)e^{-iH_5}(z_5) e^{-i(1-\epsilon)
H_5}(\zeta)e^{i(1-\epsilon) {\tilde H}_5}(\bar\zeta)\right\rangle_s\cr +& \left\langle
\sigma_{-+}^{4,1-\epsilon}(\zeta,\bar\zeta)\partial X^4(z_4) \right\rangle\left\langle
e^{iH_4/2}(x)e^{iH_4/2}(y)e^{-iH_4}(z_4) e^{-i\epsilon H_4}(\zeta) e^{i\epsilon {\tilde
H}_4}(\bar\zeta)\right\rangle_s\times \cr & \left\langle
\sigma_{+-}^{5,\epsilon}(\zeta,\bar\zeta)\partial X^5(z_5)\right\rangle \left\langle
e^{iH_5/2}(x)e^{iH_5/2}(y)e^{-iH_5}(z_5) e^{i\epsilon H_5}(\zeta)e^{-i\epsilon {\tilde
H}_5}(\bar\zeta)\right\rangle_s \cr \sim\ &\theta_{s,h_4} \left({1\over
2}(x+y)-z_4+(1-\epsilon)(\zeta-\bar\zeta)\right) \theta_{s,h_5} \left({1\over
2}(x+y)-z_5-(1-\epsilon)(\zeta-\bar\zeta)\right)\times\cr &\left[
{E(z_4,\bar\zeta)E(z_5,\zeta)\over E(z_4,\zeta)E(z_5,\bar\zeta)} \right]^{1-\epsilon}{1\over
E(\zeta,\bar\zeta)^{2(1-\epsilon)^2}} \left\langle
\sigma_{+-}^{4,1-\epsilon}(\zeta,\bar\zeta)\partial X^4(z_4) \right\rangle \left\langle
\sigma_{-+}^{5,\epsilon}(\zeta,\bar\zeta)\partial X^5(z_5)
    \right\rangle
\cr +\, &\theta_{s,h_4}
\left({1\over 2}(x+y)-z_4-\epsilon(\zeta-\bar\zeta)\right)
\theta_{s,h_5}
\left({1\over 2}(x+y)-z_5+\epsilon(\zeta-\bar\zeta)\right)\times\cr
&\left[ {E(z_4,\zeta)E(z_5,\bar\zeta)\over E(z_4,\bar\zeta)E(z_5,\zeta)}
\right]^{\epsilon}{1\over E(\zeta,\bar\zeta)^{2\epsilon^2}}
\left\langle \sigma_{-+}^{4,1-\epsilon}(\zeta,\bar\zeta)\partial X^4(z_4)
\right\rangle
\left\langle  \sigma_{+-}^{5,\epsilon}(\zeta,\bar\zeta)\partial X^5(z_5)
    \right\rangle
\, ,}} Note that due to the $H_{4,5}$ internal charge conservation, there are no correlators
involving $\sigma_{++}$ twist  fields. The spin structure sum can be performed using the
same gauge condition \gauged, with the result that the factor $K$ of Eq.\fact\ is replaced
by: \eqn\Kthreesum{\eqalign{K(z_4,z_5)&\to \int d^2\zeta\;
K_{\epsilon}(z_4,z_5,\zeta,\bar\zeta)=\cr &\int d^2\zeta\;\theta_{h_4^{-1}}\left[{0\atop
{1\over 2}\vec 1}\right] \left(z_4-\epsilon(\zeta-\bar\zeta)-\Delta\right)
\theta_{h_5^{-1}}\left[{0\atop {1\over 2}\vec 1}\right]
\left(z_5+\epsilon(\zeta-\bar\zeta)-\Delta\right)\cr &\times\left[
{E(z_4,\zeta)E(z_5,\bar\zeta)\over E(z_4,\bar\zeta)E(z_5,\zeta)} \right]^{\epsilon}{1\over
E(\zeta,\bar\zeta)^{2\epsilon^2}} \left\langle
\sigma_{-+}^{4,1-\epsilon}(\zeta,\bar\zeta)\partial X^4(z_4) \right\rangle \left\langle
\sigma_{+-}^{5,\epsilon}(\zeta,\bar\zeta)\partial X^5(z_5)
    \right\rangle\cr &\cr
 & \qquad + (4\leftrightarrow 5, \epsilon\leftrightarrow 1-\epsilon)
\, ,}} The above expression is no longer symmetric in $z_4\leftrightarrow z_5$ (except for
$\epsilon=1/2$ $\ZZ_2$ twist) therefore it can survive the antisymmetrization. Next, we will
estimate the magnitudes of the antisymmetric part and of the corresponding gaugino mass
term.

We are interested in the limit of large Scherk-Schwarz radius $R$, {\it i.e.} the limit of
low gravitino mass $m_{3/2}\sim 1/R$ in string units. As explained in \at, in the
$R\to\infty$ limit, the dominant contribution to the gaugino mass comes from the
$\tau_2\to\infty$ region of the moduli space describing the handle degeneration limit. In
this limit, $\Sigma$ degenerates into a disk with two punctures left-over from the handle.
As usual, the disk can be mapped into the upper half of the complex plane. Then the distance
between the punctures is controlled by the remaining (real) modulus $l$ of $\Sigma$. The
gaugino mass has the form \at: 
\eqn\mgz{m_{1/2}\sim g^4\int d\tau_2\int { dl\over (e^{2\pi
l}-1)} {\Gamma\over (\tau_2-l)^{3/2}}\sum_{m} {mR^2\over (\tau_2+l)^2}\, \exp({\displaystyle
-{m^2\pi R^2\over \tau_2-l}})\, . } 
Here, the sum is over the winding modes on the
Scherk-Schwarz circle, and the factor $(\tau_2-l)^{-3/2}$  includes $(\tau_2-l)^{-1}$ from
the corresponding zero modes and  $(\tau_2-l)^{-1/2}$ from the partition function. The
factor $(\tau_2+l)^{-2}$ originates from the the non-compact zero modes (four-dimensional
momenta). The factor $(e^{2\pi l}-1)^{-1}$ is the combined effect of the integration measure
and of non-zero mode determinants. We denote by $\Gamma$ any additional moduli-dependence
that may appear as a result of antisymmetrizing $K_{\epsilon}(z_4,z_5,\zeta,\bar\zeta)$,
Eq.\Kthreesum. Note that the integral over the modulus $l$ is dominated by the $l\to 0$
region. In fact, if $\Gamma$ does not vanish in this limit,  the logarithmic divergence may
give rise to the additional $\tau_2$ dependence due to the cutoff $l>e^{-\pi\tau_2}$ \at.
Thus the key question is the $\tau_2\to\infty$, $l\to 0$ behavior of $\Gamma$  -- its
``double degeneration limit'' {\it i.e.} the limit of two coalescing punctures on the disk.

In order to extract the leading $l\to 0$ behavior of
$K_{\epsilon}(z_4,z_5,\zeta,\bar\zeta)$, we can set $\zeta=\bar\zeta =0$ inside the
arguments of the theta functions in Eq.\Kthreesum. Furthermore, the twist correlators are
evaluated on the disk, and they are completely determined by the $SL(2,R)$ covariance:
\eqn\tcorr{\left\langle \sigma_{+-}^{\epsilon}(\zeta,\bar\zeta)\partial
X(z)\right\rangle={(\zeta-\bar\zeta)^{\epsilon^2}\over (z-\zeta)^2}\left({z-\zeta \over
z-\bar\zeta}\right)^{\epsilon }, \qquad \left\langle
\sigma_{-+}^{(1-\epsilon)}(\zeta,\bar\zeta)\partial
X(z)\right\rangle={(\zeta-\bar\zeta)^{\epsilon^2}\over (z-\bar\zeta)^2} \left({z-\bar\zeta
\over z-\zeta}\right)^{\epsilon}     } 
After taking the corresponding limit of the
prime-forms, $E(w_1,w_2)\to (w_1-w_2)^{-1}$, we obtain
\eqn\Kthreeb{K_{\epsilon}(z_4,z_5,\zeta,\bar\zeta)\to\theta_{h_4^{-1}}\left[{0\atop {1\over
2}\vec 1}\right] \left(z_4-\Delta\right) \theta_{h_5^{-1}}\left[{0\atop {1\over 2}\vec
1}\right] \left(z_5-\Delta\right){1\over (z_4-\bar\zeta)^2 (z_5-\zeta)^2}  +
(4\leftrightarrow 5)} 
Note that the $\epsilon$-dependence has disappeared in this limit. The
above function vanishes upon antisymmetrization, thus $\Gamma={\cal O}(l)$ and the
$l$-integral in \mgz\ converges at $l=0$. The dominant $\tau_2\to\infty$ region yields
\eqn\mgy{m_{1/2}\sim g^4\int {d\tau_2\over \tau_2^{7/2}}\sum_{m} mR^2\exp({\displaystyle
-{m^2\pi R^2\over \tau_2}})\, .} 
After rescaling $\tau_2\to \tau_2 R^2$, the above
expression yields\foot{This result corrects \at, where a different conclusion has been
reached without appropriate analysis of the twisted correlators.} 
\eqn\mgw{m_{1/2} \sim{g^4\over R^3} \;\sim\; g^4m_{3/2}^3\, .}

The mass \mgw\ can be understood within the effective field theory by looking at a generic
one-loop graph involving a gravitational exchange. Each vertex brings one power of the Planck
mass $M_P$ in the denominator and is quadratically divergent in the ultraviolet, thus $m_{1/2}\sim
m_{3/2} \Lambda^2_{UV}/M_P^2$, where $\Lambda_{UV}$ is the ultraviolet cutoff. This cutoff
is should be of order of the supersymmetry breaking scale \cutoff,\foot{We are grateful to Savas
Dimopoulos and to Mary K. Gaillard for insisting on this point.} $\Lambda_{UV}\sim m_{3/2}$,
hence $m_{1/2}\sim m_{3/2}^3/ M_P^2$ \mgaugino. The result \mgw\ confirms this expectation.

To summarize, the mediation of supersymmetry breaking from closed to open string sectors
provides a superstring realization of the so-called gravitational mediation \pran. 
Another type of mediation, the anomaly mediation \rs\ discussed in the beginning and in \at, 
is absent, at least at the leading ${\cal O}(g_s)\sim {\cal O}(g^2)$ order. 
Moreover, it is also absent at the ${\cal O}(g^4)$ order, as follows from
the result \mgw.
A new type of non-gravitational mediation between open string sectors has 
been recently discussed in \antmass.
\bigskip
\bigskip

We are grateful to Savas Dimopoulos and to Mary K. Gaillard for enlightening discussions and
correspondence, and to Narain for his continuous help.
This work  is supported in part by the European Commission under the RTN contracts
MRTN-CT-2004-503369 and MEXT-CT-2003-509661, in part by the INTAS contract 03-51-6346, and
in part by the CNRS PICS \# 2530. The research of T.R.T.\ is supported in part by the U.S.
National Science Foundation Grant PHY-0242834. Any opinions, findings, and conclusions or
recommendations expressed in this material are those of the authors and do not necessarily
reflect the views of the National Science Foundation.

    \listrefs

\bye
\end